\documentclass[12pt]{article}

\usepackage{amsmath,amssymb,amsfonts}   
\usepackage{graphicx,bm}

\usepackage{geometry}

    \newgeometry{vmargin={20mm,30mm}, hmargin={25mm,25mm}}  
    
    \usepackage[margin=1cm]{caption}

\newcommand{\calU}{{\mathcal U}}
\newcommand{\calN}{{\mathcal N}}

\newcommand{\calG}{{\mathcal G}}
\newcommand{\R}{{\mathbb R}}

\newcommand{\x}{\mathbf{x}}
\newcommand{\y}{\mathbf{y}}
\newcommand{\e}{{\mathrm e}}

\newcommand{\n}{\mathbf n}

\begin{document}

\title{Asymptotic analysis of particle cluster formation in the presence of anchoring sites}
\author{ \em
Paul. C. Bressloff, \\ Department of Mathematics, Imperial College London, \\
London SW7 2AZ, UK}

\maketitle

\abstract{
The aggregation or clustering of proteins and other macromolecules plays an important role in the formation of large-scale molecular assemblies within cell membranes. Examples of such assemblies include lipid rafts, and postsynaptic domains (PSDs) at excitatory and inhibitory synapses in neurons. PSDs are rich in scaffolding proteins that can transiently trap transmembrane neurotransmitter receptors, thus localizing them at specific spatial positions. Hence, PSDs play a key role in determining the strength of synaptic connections and their regulation during learning and memory. Recently, a two-dimensional (2D) diffusion-mediated aggregation model of PSD formation has been developed in which the spatial locations of the clusters are determined by a set of fixed anchoring sites. The system is kept out of equilibrium by the recycling of particles between the cell membrane and interior. This results in a stationary distribution consisting of multiple clusters, whose average size can be determined using an effective mean-field description of the particle concentration around each anchored cluster. In this paper, we derive corrections to mean-field theory using asymptotic methods, assuming that the clusters are well separated and much smaller than the membrane surface. The analysis proceeds by partitioning the membrane into a set of inner regions around each cluster, and an outer region where mean-field interactions occur. Asymptotically matching the inner and outer stationary solutions, generates an asymptotic expansion of the particle concentration that sums over all logarithmic terms involving the small parameter $\nu=-1/\ln \epsilon$, where $\epsilon$ specifies the relative size of the clusters. This is then used to derive an expression for the radius of each cluster, which includes higher-order corrections to  mean-field theory that depend on the positions of the clusters and the boundary of the domain. Finally, motivated by a recent study of light-activated protein oligomerization in cells, we develop the analogous theory for cluster formation in a three-dimensional (3D) domain.}
%%%%%%%%%%%%%%%%%%%%%%%%%%%

\section{Introduction}

The aggregation or clustering of proteins and other macromolecules plays an important role in the formation of large-scale molecular assemblies in cells. In many cases such assemblies are associated with cellular membranes, including the clustering of cell-cell adhesion proteins in epithelia \cite{Quang13} and lipid raft formation \cite{Turner05,Turner18}.
Another notable example is the formation of postsynaptic domains (PSDs) at excitatory and inhibitory synapses in neurons. PSDs are rich in scaffolding proteins that can transiently trap transmembrane neurotransmitter receptors, thus localizing them at specific spatial positions, in particular, at sites apposed to active zones in presynaptic domains where neurotransmitters are released. PSDs thus play a crucial role in determining the effective strength of synaptic connections between cells  \cite{Meier01,Borgdorff02,Dahan03,Choquet03,Groc04,Choquet13,Salvatico15,Choquet18,Triller23}. Advances in single-particle tracking and imaging methods have shown that PSDs are highly dynamic structures whose constituent molecular components are subject to continuous turnover. For example, scaffolding protein-receptor complexes can diffuse laterally within the cell membrane. A surface complex may also be
internalized via endocytosis and stored within an intracellular compartment, where it is either recycled to the surface via recycling endosomes and exocytosis, or sorted for degradation by late endosomes and lysosomes.

A number of models have explored the combined effects of diffusion-trapping and recycling on the number of excitatory $\alpha$-amino-3-hydroxy-5-methyl-4-isoxazolepropionic (AMPA) receptors within dendritic spines \cite{Earnshaw06,Holcman06,Bressloff07,Earnshaw08,Thoumine12,Triesch18}. However, these diffusion-trapping models typically assume that the number of trapping sites or ``slots'' within a given PSD is fixed. In order to understand the formation and stabilization of PSDs, it is necessary to consider the slower dynamics of scaffolding protein-receptor complexes. Several modeling studies have analyzed the joint localization of gephyrin scaffolding proteins and glycine receptors at inhibitory synapses \cite{Sekimoto09,Hasel11,Hasel15}, showing how stable PSDs could arise dynamically via a nonequilibrium Turing mechanism. An alternative approach is to consider PSD formation in terms of diffusion-mediated particle aggregation or coalescence within the cell membrane \cite{Ranft17,Hakim20}. The fusion of smaller diffusing clusters and particles is modeled according to Smoluchowski coagulation equations, and the system is kept out of equilibrium by the recycling of particles between the cell membrane and interior \cite{Foret12}. This allows for the formation and maintenance of a stationary distribution consisting of multiple clusters. Recently, the PSD aggregation model has been extended to include fixed anchoring sites that trap clusters at specific positions within the membrane, consistent with the alignment of PSDs and presynaptic active zone \cite{Ranft23}.
The effects of the anchoring sites were analyzed using a mean field description of the steady-state particle concentration around a single cluster. This yielded an expression for the mean cluster size as function of model parameters such as the density of anchoring sites. 

In this paper, we extend the mean field analysis of cluster formation in the presence of anchoring sites, in order to take into account diffusion-mediated interactions between the anchored clusters. We assume that the clusters are well separated and much smaller than the membrane surface, so that we can apply the theory of diffusion in domains with small holes or exclusions (strongly localized perturbations) \cite{Ward93,Ward93a,Straube07,Bressloff08,Coombs09,Cheviakov11,Chevalier11,Ward15,Lindsay16,Bressloff21a}. 
After presenting the basic 2D aggregation model and its mean field formulation in Sect. 2, we develop the asymptotic analysis in Sect. 3. We proceed by partitioning the region outside the clusters into a set of inner regions around each cluster together with an outer region where mean-field interactions occur. The inner solutions depend on the small parameter $\nu=-1/\ln \epsilon $, where $\epsilon$ specifies the relative size of each cluster. The presence of the small parameter $\nu$ rather than $\epsilon$ is a common feature of strongly localized perturbations in 2D domains. Since $\nu \rightarrow 0$ much more slowly than $\epsilon\rightarrow 0$, it is necessary to sum over the logarithmic terms non-perturbatively in order to obtain $O(\epsilon)$ accuracy. This can be achieved by matching the inner and outer solutions using Green's functions \cite{Ward93}, which is equivalent to calculating the asymptotic solution for all terms of $O(\nu^k)$ for any $k$. We thus derive an expression for the steady-state particle concentration outside the clusters that depends on the positions of the various anchor points and the boundary of the 2D domain. The steady-state cluster sizes are then obtained by calculating the total flux of particles at the surface of each cluster. 
 
 In Sect. 4 we consider an analogous 3D model of clustering in the presence of anchoring sites. One major difference from the 2D model is that the recycling of particles can no longer be interpreted in terms of membrane exo/endocytosis. Instead, we assume that particles exist in either an active or inactive state, and can only form clusters in the active state. One motivation for the 3D model is a recent study of the optogenetic protein CRY2olig, which oligomerizes (forms small clusters) in the presence of blue light \cite{Coppey21}. (In this particular study, the authors explore both theoretically and experimentally the effects of obstacles on the formation of large 3D protein clusters via the diffusion-limited aggregation of oligomers.) We show how the asymptotic analysis of the 3D model is significantly different from the 2D case, due to the distinct singularity structure of the 3D Green's function. 
 
The asymptotic analysis presented in this paper complements our previous work on the active suppression of Ostwald ripening \cite{Bressloff20a,Bressloff20b}. Ostwald ripening describes the coarsening of droplets during late-stage liquid-liquid phase separation via spinodal decomposition \cite{Lif61,Wagner61}, and is distinct from the coarsening process underlying aggregation models. In classical Ostwald ripening, an emulsion of polydisperse droplets transitions to a single condensate in thermodynamic equilibrium with a surrounding dilute phase. However, a key feature of biological liquid-liquid phase separation is the co-existence of multiple condensates over significant time scales \cite{Hyman14,Brangwynne15,Berry18,Mittag22}. Various hypotheses have been given to account for the suppression of Ostwald ripening, including actively driven chemical reactions that maintain the out-of-equilibrium switching of proteins between soluble and phase separating forms \cite{Zwicker15,Wurtz18,Lee19,Weber19}. In our previous work we used asymptotic analysis to study the existence of multiple condensate solutions for this type of model.

\setcounter{equation}{0} 
\section{Cluster formation model and the mean field approximation}

\begin{figure}[t!]
\centering
\includegraphics[width=14cm]{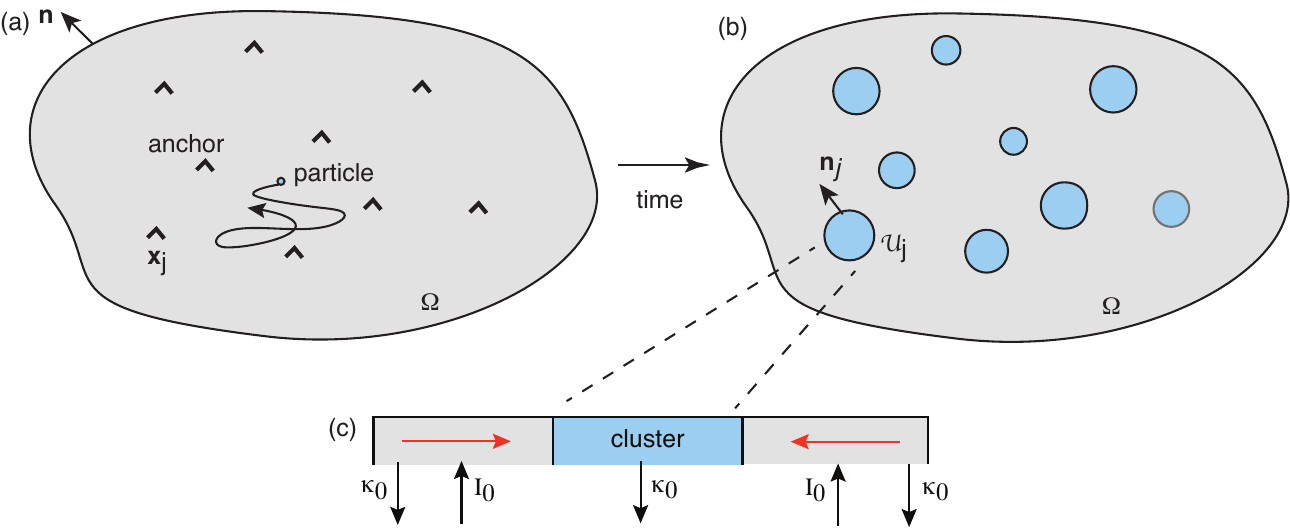} 
\caption{\small 2D model of diffusion-based protein cluster formation in the presence of anchoring cites and particle recycling. (a) A set of $N$ anchoring sites at positions $\x_j$, $j=1,\ldots,N$, in a bounded domain $\Omega$. (b) Diffusing particles accumulate at the anchoring sites resulting in the formation of particle aggregates or clusters $\calU_j$. (c) The clusters are dynamically maintained by a combination of lateral diffusion outside the clusters and particle recycling.}
\label{fig1}
\end{figure}

Consider a bounded domain $\Omega \subset \R^2$ containing a set of $N$ anchoring sites $\x_j \in \Omega$, $j=1,\ldots,N$, see Fig. \ref{fig1}(a). Suppose that at time $t$, $t\geq 0$, there exists a circularly symmetric cluster $\calU_j(t)$ of radius $r_j(t)$ at the $j$th site, see Fig. \ref{fig1}(b). That is, $\calU_j(t)=\{\x\in \Omega,\ |\x-\x_j|\leq r_j(t)\}$. Let $c(\x,t)$, $\x \in \Omega$, denote the concentration of freely diffusing particles (monomers), which evolves according to the diffusion equation
\begin{subequations} 
\label{cluster}
\begin{align}
	\frac{\partial c(\x,t)}{\partial t} &= D{\bm \nabla}^2 c(\x,t)-\kappa_0 c(\x,t)+I_0,\nonumber  \\ & \x\in \Omega \backslash \cup_{j=1}^N  \calU_j(t),\\
	D{\bm \nabla} c(\x,t) \cdot \n&=0,\  \x\in\partial \Omega ,\\
	c(\x,t)&=0,\quad \x\in \partial \calU_j(t).	\end{align}
Here $\n$ is the outward unit normal to the surface $\partial \Omega$, the constant $\kappa_0$ denotes the recycling or turnover rate of individual particles outside a cluster, and $I_0$ is the re-injection flux of recycled particles, see Fig. \ref{fig1}(c). It remains to specify the dynamics of cluster growth. Let $\calN_j(t)$ denote the total number of particles within the $j$th cluster at time $t$. It follows from particle conservation that 
\begin{align}
\frac{d}{dt}\calN_j(t) &= D\int_{\partial \calU_j(t)}{\bm \nabla} c(\x,t) \cdot \n_jd\x-\kappa_0 \calN_j(t),
\end{align}
\end{subequations} 
where ${\bf n}_j$ is the outward unit normal to the surface $\partial \calU_j$. In the case of a uniform density $u_0$ of particles within a cluster, the number of particles is $\calN_j(t)=|\calU_j(t)|u_0$. For concreteness, we assume that the rate of recycling within a cluster is also $\kappa_0$, and that no particles are re-injected directly into a cluster (as appears to hold for PSD formation). However, one could consider a more general model, in which the recycling rates within and outside clusters differ.

In this paper we are interested in calculating the steady-state solution. We proceed by solving the  steady-state version of Eqs. (\ref{cluster}a-c) in terms of the unknown steady-state radii $r_j$, $j=1,\ldots,N$:
\begin{subequations} 
\label{SScluster}
\begin{align}
	D{\bm \nabla}^2 c(\x)-\kappa_0 c(\x)+I_0&=0,\  \x\in \Omega \backslash \cup_{j=1}^N  \calU_j,\\
	D{\bm \nabla} c(\x) \cdot \n&=0,\  \x\in\partial \Omega ,\\
	c(\x)&=0,\quad \x\in \partial \calU_j.	\end{align}
	\end{subequations} 
 The radii are then determined self-consistently by imposing the steady-state version of Eq. (\ref{cluster}d):
 \begin{equation}
 \label{SScon}
 J_j:=D\int_{\partial \calU_j}{\bm \nabla} c(\x) \cdot \n_jd\x=\kappa_0 u_0 |\calU_j|
 \end{equation}
 for $j=1,\ldots,N$, where $J_j$ is the total flux into the cluster. 
 
 For the moment, suppose that each cluster is treated as a point source/sink. That is, we replace Eqs. (\ref{SScluster}) by
\begin{align}
\label{diff}
D{\bm \nabla}^2 c(\x)-\kappa_0 c(\x)+I_0-\sum_{j=1}^N J_j \delta(\x-\x_j)&=0,
\end{align} 
together with the boundary conditions $c(\x_j)=0$, $j=1,\ldots,N$. 
Consider the Neumann Green's function $G(\x,\y)$ for the modified Helmholtz equation, which is uniquely defined by
\begin{subequations}
\label{G1}
\begin{align}
D{\bm \nabla}^2 G-\kappa_0 G&=-\delta(\x-\x'),\quad \x,\x' \in \Omega,\\ {\bm \nabla G}\cdot \n & =0 \mbox{ on } \partial \Omega
\end{align}
\end{subequations}
for fixed $\x'$. Note that $G $ can be decomposed as
\begin{equation}
G(\x,\x')=-\frac{ \ln |\x-\x'|}{2\pi D}+R(\x,\x'),
\end{equation}
where $R$ is the regular (non-singular) part of the Green's function. It follows that the formal solution of Eq. (\ref{diff}) is given by
\begin{align}
c(\x)&=\int_{\Omega} G(\x,\x')\left [I_0-\sum_{k=1}^NJ_j \delta(\x'-\x_k)\right ]d\x'\nonumber \\
&=c_0-\sum_{k=1}^NJ_kG(\x,\x_k),
\end{align}
where $c_0=I_0/\kappa_0$ and we have used the fact that $\int_{\Omega}G(\x,\x')d\x'=\kappa_0^{-1}$. In principle, the unknown fluxes $J_j$ could now be obtained by imposing the conditions $c(\x_j)=0$, which would yield the matrix equation
$c_0=\sum_{k=1}^NG(\x_j,\x_k)J_k$ for $j=1,\ldots,N$. 
However, the Green's function $G(\x,\x')$ has a logarithmic singularity in the limit $\x\rightarrow \x'$, so that the diagonal elements $G(\x_j,\x_j)=\infty$.

\begin{figure}[b!]
\centering
\includegraphics[width=10cm]{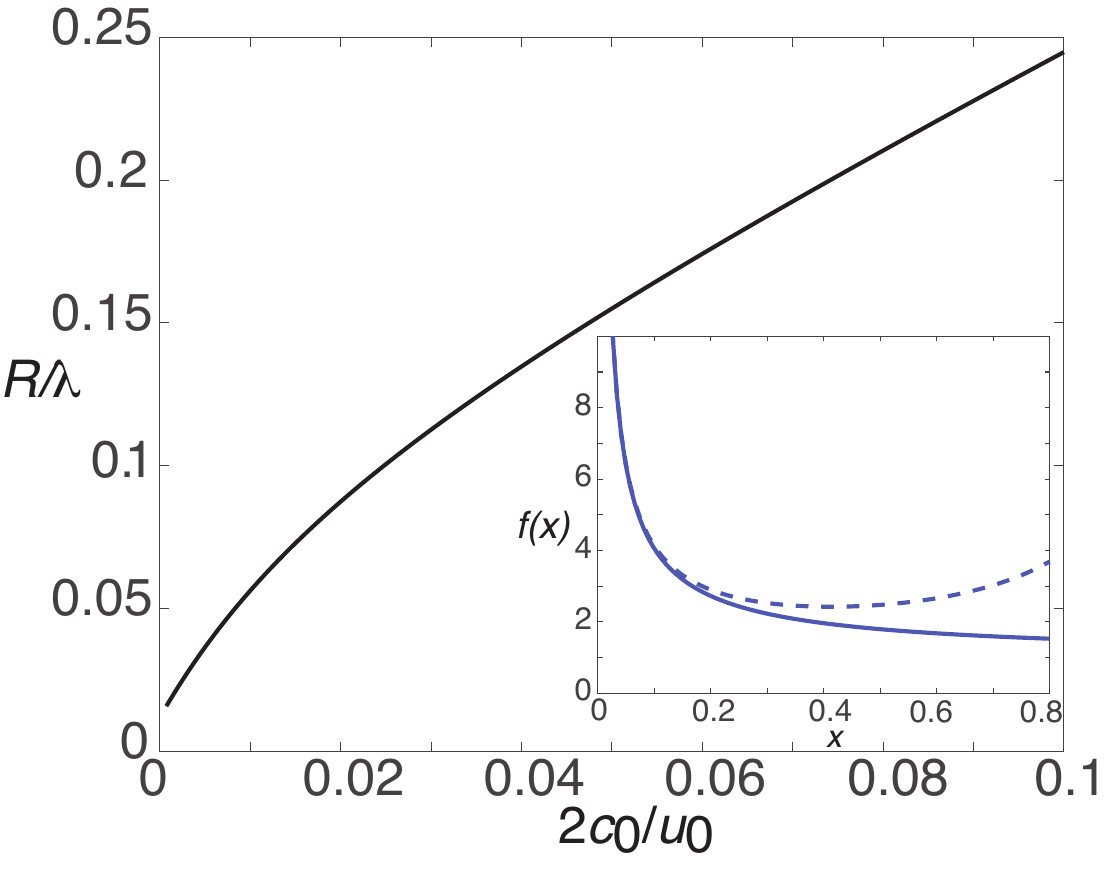} 
\caption{\small Mean-field approximation in the dilute cluster regime. Plot of non-dimensionalized mean cluster radius $R/\lambda$ as a function of $2c_0/u_0$, where $c_0$ is the stationary particle concentration away from clusters and $u_0$ is the particle concentration within a cluster. {\em Inset}: Plot of $f(x)=K_1(x)/K_0(x)$ as a function of $x$ (solid curve). The asymptotic approximation of $f(x)$ obtained from Eq. (\ref{Kap}) is shown as the dashed curve.}
\label{fig2}
\end{figure}

An alternative approach is to coarse-grain the system by treating each cluster as if it is in a ``sea'' of uniformly distributed background clusters \cite{Ranft23}. Assume that all of the clusters have the same radius $R$ and that the background cluster density is $\phi_0$. Ignoring the effects of the boundary $\partial \Omega$, we can take the concentration $c$ around a single cluster to be circularly symmetric. The coarse-grained version of Eq. (\ref{diff}) for any individual cluster is then 
\begin{align}
\label{mft}
D{\bm \nabla}^2 c(r)-\kappa_0 c(r)+I_0-\phi_0 J(R)&=0,\ r >R,
\end{align} 
with $c(R)=0$ and the flux $J(R)$ into the surface of each cluster determined self-consistently from the equation
\begin{equation}
\label{coJ}
J(R)=2\pi R D\partial_rc(R).
\end{equation}
This 2D mean-field approximation was analyzed in Ref. \cite{Ranft23}. First, the solution of Eq. (\ref{mft}) is given by
\begin{equation}
c(r)=\left (c_0-\frac{\phi_0J(R)}{\kappa_0}\right )\left (1-\frac{K_0(r/\lambda)}{K_0(R/\lambda)}\right ),
\end{equation}
where $K_{\nu}$ is a modified Bessel function of the second kind, $c_0=I_0/\kappa_0$, and $\lambda=\sqrt{D/\kappa_0}$. Substituting the solution into Eq. (\ref{coJ}) yields a self consistency equation for the flux $J(R)$:
\begin{align}
\label{mftJ}
J(R)=\frac{2\pi RD}{\lambda} \left (c_0-\frac{\phi_0J(R)}{\kappa_0}\right )\frac{K_1(R/\lambda)}{K_0(R/\lambda)}.
\end{align}
We have used the Bessel identity $K_0'(x)=-K_1(x)$. Eq. (\ref{mftJ}) can be rewritten as an implicit equation for the cluster radius $R$ by using Eq. (\ref{SScon}), since $J(R)=\pi R^2 \kappa_0 u_0$ \cite{Ranft23}:
\begin{align}
\label{mftR}
\frac{R}{\lambda}=2 \left (\frac{c_0}{u_0}-\pi\left ( \frac{R}{\lambda}\right )^2 \phi_0\lambda^2\right )\frac{K_1(R/\lambda)}{K_0(R/\lambda)}.
\end{align}

The mean cluster radius $R$ can now be determined as a function of model parameters by solving Eq. (\ref{mftR}) numerically \cite{Ranft23}. In anticipation of the subsequent asymptotic analysis, suppose $\phi_0 \pi \lambda^2 \ll 1$ (dilute cluster regime), so that
\begin{equation}
\frac{R}{\lambda}\approx \frac{2c_0}{u_0}f(R/\lambda),\quad f(R/\lambda)=\frac{K_1(R/\lambda)}{K_0(R/\lambda)}.
\end{equation}
The solution for $R/\lambda$ is plotted as a function of the parameter $2c_0/u_0$ in Fig. \ref{fig2}. It can be seen that if $2c_0/u_0$ is sufficiently small $\Gamma$ (as assumed in Ref. \cite{Ranft23}), then the mean cluster radius is relatively small. Therefore, we can exploit the small-$z$ expansions (see inset of Fig. \ref{fig2})
\begin{equation}
\label{Kap}
K_0(z)\sim -\ln (z/2) -\gamma_c,\quad K_1(z)\sim \frac{1}{z},
\end{equation}
where $\gamma_c \approx 0.5772$ is Euler's gamma constant. Setting $\epsilon^2=2c_0/u_0$ and $R/\lambda =\epsilon \rho/\lambda  $ with $\rho/\lambda =O(1)$,
 we thus find that 
\begin{align}
\left (\frac{ \rho}{\lambda}\right )^2 &\sim  \frac{1}{-\ln \epsilon  +\ln 2-\ln \rho/\lambda -\gamma_c}\nonumber \\
&= \frac{\nu}{1+\nu \left(\ln 2-\ln \rho/\lambda  - \gamma_c\right)},
\label{mftX}
\end{align}
where $\nu=-1/\ln \epsilon$. As we highlighted in the introduction, the non-perturbative dependence on the small parameter $\nu$ is a common feature of strongly localized perturbations in 2D domains \cite{Ward93}. In Sect. III we determine corrections to the mean-field result (\ref{mftX}) by solving the full steady-state Eqs. (\ref{SScluster}) in the small cluster limit. These corrections take into account diffusion-mediated interactions between the clusters as well as the effects of the boundary $\partial \Omega$.

\setcounter{equation}{0}
 \section{Asymptotic analysis in 2D}
In order to solve the BVP given by Eqs. (\ref{SScluster}a-c),  we fix the length scale by setting $L:=|\Omega|^{1/2}=1$. We then take the anchoring sites to be well-separated in the sense that $|\x_i-\x_j| =O(1)$ for all $j\neq i$ and $\min_{\y}\{|\x_j -\y|,\y \in \partial \Omega \} =O(1)$, $j=1,\ldots,N$. Finally, as in the mean-field approximation (\ref{mftX}), we assume that $u_0=2c_0/\epsilon^2$ and write the steady-state radius of the $j$th cluster as $r_j=\epsilon  \rho_j$ with $0<\epsilon \ll 1$ and $\rho_j=O(1)$. Under these various conditions, we can use a combination of matched asymptotics and Green's function methods along analogous lines to the classical narrow capture problem. More specifically, we construct an inner or local solution
valid in an $O(\epsilon)$ neighborhood of each cluster, and then match to an outer or global solution that is valid away from each neighborhood. 

 First, consider the inner solution around the $j$th cluster,
\[C_j(\y)=c(\x_j+\epsilon^{1/2} \y),\quad \y=\epsilon^{-1}(\x-\x_j),\]
where we have introduced stretched coordinates and replaced the domain $\Omega $ by $\R^2$. It follows that
\begin{subequations}
\begin{align}
{\bm \nabla}_{\y}^2 C_j(\y)&=0 \mbox{ for } |\y| >\rho_j, \\
C_j(\y)&= 0\mbox{ for } |\y|=\rho_j,
\end{align}
\end{subequations}
which can be expressed in polar coordinates as
\begin{align}
\label{2Dsymin}
\frac{1}{\rho}\frac{d}{d\rho}\rho\frac{dC_j}{d\rho}&=0,\, \rho_j< \rho<\infty,\ C_j(\rho_j)=0.
\end{align}
The inner solution takes the form
\begin{equation}
\label{inner0}
C_{j}(\rho)=  A_j  \ln( \rho /\rho_j),\quad \rho\geq \rho_j,
\end{equation}
where $A_j$ is some undetermined coefficient.
The corresponding solution in the original coordinates is
\begin{equation}
\label{inner}
C_j(\x)= A_j \ln( |\x-\x_j |/\epsilon \rho_j ).
\end{equation}
The coefficients $A_j $, $j=1,\ldots,N$, can be determined by matching the inner solutions with the corresponding outer solution (see below). 

The outer solution is obtained by treating each cluster as a point source/sink localized at the anchor sites. The corresponding diffusion equation takes the form
\begin{subequations}
\label{outer0}
\begin{align}
D{\bm \nabla}^2 c(\x)-\kappa_0c(\x)+I_0=&0,\quad \x\in \Omega\backslash \{\x_1,\ldots,\x_N\}, \\ {\bm \nabla}c(\x)\cdot \n&=0,\quad \x \in \partial \Omega,
\end{align}
together with the matching condition
\begin{equation}
\label{match}
c(\x)\sim  \frac{A_j}{\nu}+  A_j \ln [|\x-\x_j|/\rho_j], \quad \nu=-\frac{1}{\ln \epsilon }
\end{equation}
\end{subequations}
as $\x\rightarrow \x_j$. The next step is to introduce the ansatz 
\begin{equation}
\label{outer}
c(\x)\sim  c_0-2\pi   D \sum_{j=1}^NA_j G(\x,\x_j)
\end{equation}
for $\x \notin \{\x_j,\, j=1,\ldots,N\}$
and some constant $c_0$. Here $G$ is the modified Helmholtz Green's function satisfying Eqs. (\ref{G1}).
Observe that for $\x \notin \{\x_j,\, j=1,\ldots,N\} $,
\begin{align}
&D{\bm \nabla}^2 c(\x) -\kappa_0c(\x)\\
&\sim -\kappa_0 c_0 -2\pi D \sum_{i=1}^NA_i [D{\bm \nabla}^2-\kappa_0] G(\x,\x_i)=-\kappa_0 c_0.\nonumber 
\end{align}
Hence, the outer solution satisfies Eq. (\ref{outer0}) provided that $c_0=I_0/\kappa_0$. In addition, integrating Eq. (\ref{G1}a) with respect to $\x\in \Omega$ shows that $\int_{\Omega}G(\x,\x')d\x=1/\kappa_0$. Applying this result to Eq. (\ref{outer}) implies that
\begin{equation}
\int_{\Omega}c(\x)d\x\sim \kappa_0^{-1}\bigg (|\Omega|I_0-2\pi  D\sum_{j=1}A_j \bigg ).
\end{equation}

We now determine the unknown coefficients $A_j $, $j=1,\ldots,N$, by matching the inner and outer solutions. As $\x\rightarrow \x_j$, we have
\begin{align}
c(\x)&\rightarrow c_0 +   A_j \ln|\x-\x_j|-2\pi   D A_j R(\x_j,\x_j)\nonumber \\
&\quad -2\pi   D\sum_{i\neq j}^NA_i G (\x_j,\x_i).
\end{align}
Comparison with the asymptotic limit in equation (\ref{match}) yields the self-consistency conditions
\begin{align}
c_0
&=\left [\frac{1}{\nu}-  \ln \rho_j+2\pi  D R(\x_j,\x_j) \right ]A_j \nonumber \\
&\quad +2\pi  D \sum_{i\neq j} A_i G (\x_j,\x_i)
\label{Ai}
\end{align}
for $ j=1,\ldots,N$. In particular, this can be rewritten as a matrix equation
\begin{equation}
\label{matrix}
\sum_{i=1}^N\left (\delta_{i,j}+2\pi \nu D \calG_{ji}\right )A_i =\nu c_0,
\end{equation}
with 
\begin{align}
\calG_{jj}&=R (\x_j,\x_j)-\frac{\ln \rho_j}{2\pi D} ,\quad  \calG_{ji}=G(\x_j,\x_i),\, j\neq i.
\label{M}
\end{align}
We thus obtain the solution
\begin{equation}
\label{Ai2}
A_i =A_i(\nu)\equiv \nu c_0\sum_{j=1}^N[1+2\pi \nu D\calG]^{-1}_{ij},
\end{equation}
which is clearly non-perturbative with respect to $\nu$.

In order to calculate the coefficients $A_i(\nu)$ we need to obtain accurate numerical or analytical approximations of the Green's function for the modified Helmholtz equation and solve the matrix equation (\ref{matrix}). This particular issue has been addressed by Lindsay {\em et al.}  \cite{Lindsay16}, whose results can be applied to the current problem. An important step in the evaluation of the Green's function is to decompose $G$ as the sum of the free-space Green's function and a regular part:
\begin{align}
    G(\x,\x') = \frac{1}{2\pi D}K_0\left(|\x - \x'|/\lambda\right) + \widehat{R}(\x,\x'),
\end{align}
where $\lambda =\sqrt{D/\kappa_0}$, $K_0$ is the modified Bessel function of the second kind and $\widehat{R}$ is non-singular at $\x =\x'$.  It can be shown that for $|\x - \x'| = O(1)$ and sufficiently small $\lambda$,  the boundary contributions to $\widehat{R}$ are exponentially small. (If we ignore boundary effects then the fundamental length-scale is given by $\lambda$.) This allows us to write
\begin{align*}
    G(\x,\x') &\sim \frac{1}{2\pi D}K_0\left(|\x - \x'|/\lambda\right), \ \x \neq \x', \\
  \widehat{R}(\x',\x') &\sim \frac{1}{2\pi D}\left(\ln\lambda  + \ln 2 - \gamma_c\right).
\end{align*}
Substituting these approximations into Eq. (\ref{Ai}) implies that. 
\begin{align}
\label{AAi}
A_j(\nu)&\sim \frac{\nu c_0}{1+\nu \left(\ln 2-\ln \rho_j/\lambda  - \gamma_c\right)}\sum_{k=1}^N[{\bm K}(\nu)^{-1}]_{jk},
\end{align}
where
\begin{align}
[{\bm K}(\nu)]_{jk}=\delta_{j,k}+ \frac{\nu K_0(|\x_j-\x_k|/\lambda)[1-\delta_{j,k}]}{1+\nu \left(\ln 2-\ln \rho_j/\lambda  - \gamma_c\right )}.
\end{align}

The asymptotic analysis has yielded an inner solution around each cluster that is expressed in terms of the unknown radii $\rho_j$, $j=1,\ldots,N$. The latter are now determined self-consistently by imposing the conditions (\ref{SScon}), which in 2D reduces to the implicit equations
\begin{equation}
2\pi \rho_j  D \frac{  A_j(\nu)}{\rho_j}= \epsilon^2 \kappa_0 {u}_0 \pi \rho_j^2 ,\quad j=1,\ldots,N.
 \end{equation}
 Substituting for $A_j(\nu)$ using Eq. (\ref{AAi}) and noting that $u_0=2c_0/\epsilon^2$ yields 
 \begin{align}
 \label{rhoj}
\left (\frac{\rho_j}{\lambda }\right )^2&\sim \frac{\nu }{1+\nu \left(\ln 2-\ln \rho_j/\lambda  - \gamma_c\right)}\sum_{k=1}^N[{\bm K}(\nu)^{-1}]_{jk}.
\end{align}
The first factor on the right-hand side of Eq. (\ref{rhoj}) recovers the mean field Eq. (\ref{mftX}). The second factor is the correction to mean field theory that involves diffusion-mediated interactions between the anchor points. 

\begin{figure}[t!]
\centering
\includegraphics[width=10cm]{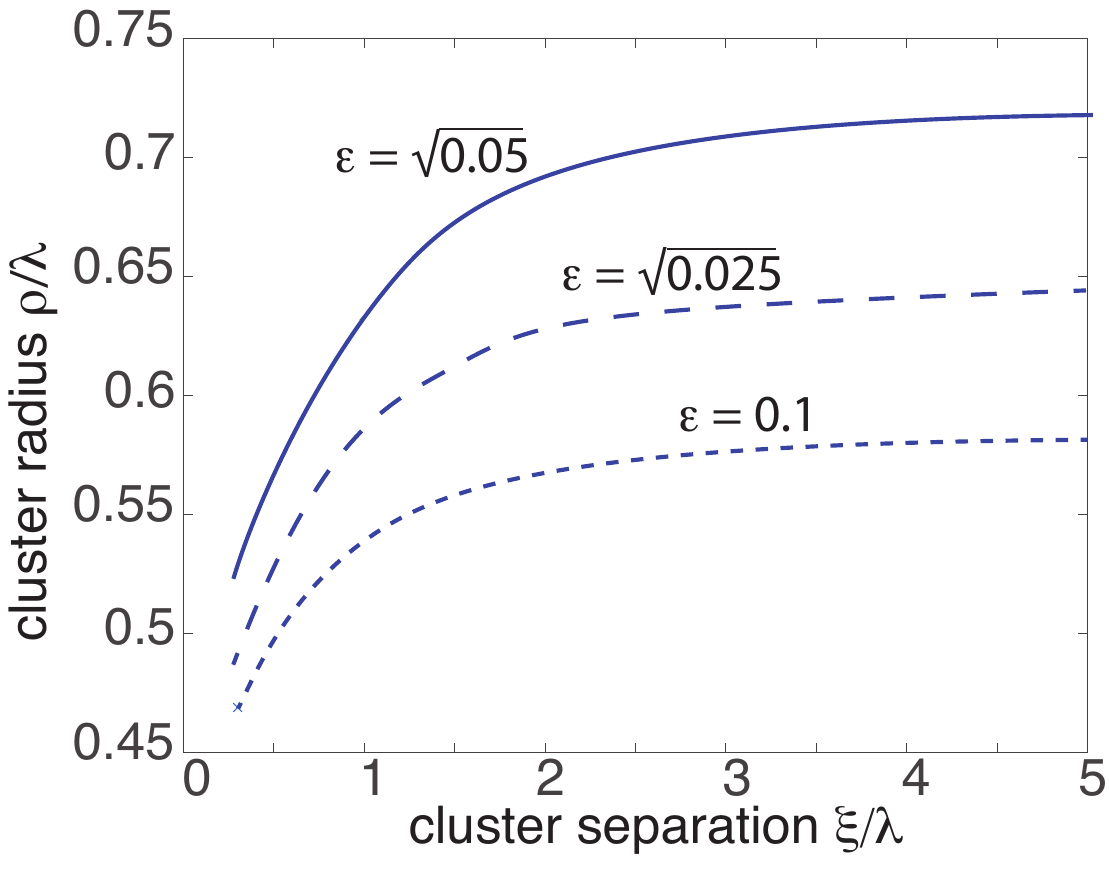} 
\caption{Pair of clusters of radius $\rho$ given by the solution to the non-perturbative Eq. (\ref{rhoj1}). The radius $\rho$ is plotted as a function of cluster separation $\xi$ for different values of $\epsilon$ with $\epsilon^2 =2c_0/u_0$. In the limit $\xi\rightarrow \infty$ we recover the mean-field results.}
\label{fig3}
\end{figure}
 
As a simple illustration consider a pair of clusters with spatial separation $\xi=|\x_1-\x_2|$. If we ignore the effects of the exterior boundary, then both clusters have the same radius $\rho_1=\rho_2=\rho$. The matrix ${\bm K}$ and its inverse are then simply
\begin{subequations}
\begin{align}
{\bm K}(\nu)&=\left (\begin{array}{cc} 1 & \Gamma(\nu,\xi)\\
\Gamma(\nu ,\xi) &1 \end{array} \right ),\\
 {\bm K}(\nu)^{-1}&=\frac{1}{1-\Gamma(\nu,\xi)^2}\left (\begin{array}{cc} 1 & -\Gamma(\nu,\xi)\\
-\Gamma(\nu ,\xi) &1 \end{array} \right ),
\end{align}
\end{subequations}
where
\begin{equation}
\Gamma(\nu,\xi)=  \frac{\nu K_0(\xi/\lambda)}{1+\nu \left(\ln 2-\ln \rho/\lambda  - \gamma_c\right )}.\end{equation}
It follows that $\rho$ is the solution to the implicit equation
\begin{align}
 \label{rhoj1}
\left (\frac{\rho}{\lambda }\right )^2&\sim \frac{\nu }{1+\nu \left(\ln 2-\ln \rho/\lambda  - \gamma_c\right)}\frac{1}{1+\Gamma(\nu,\xi)}.
\end{align}
In Fig. \ref{fig3} we plot the numerical solution of Eq. (\ref{rhoj1}) as a function of the cluster separation $\xi$ for $\epsilon^2=0.01$, $\epsilon^2=0.025$ and $\epsilon^2=0.05$. The mean-field solution is recovered in the limit $\xi\rightarrow \infty$. It can be seen that the non-perturbative corrections to mean-field theory are significant.

\begin{figure*}[t!]
\centering
\includegraphics[width=14cm]{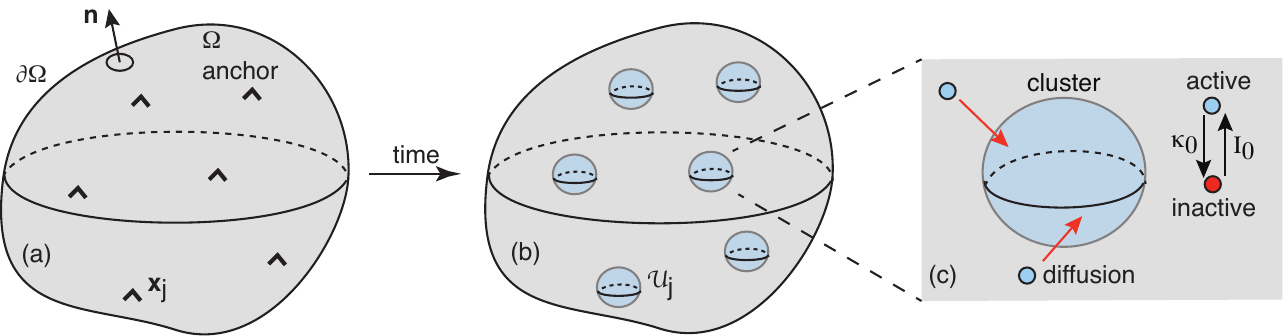} 
\caption{\small 3D model of diffusion-based protein cluster formation in the presence of anchoring cites and particle activation/deactivation. (a) A set of $N$ anchoring sites at positions $\x_j$, $j=1,\ldots,N$, in a bounded domain $\Omega$. (b) Diffusing particles in the activated state accumulate at the anchoring sites resulting in the formation of particles aggregates or clusters. (c) The clusters are dynamically maintained by a combination of diffusion and activation/deactivation.}
\label{fig4}
\end{figure*}

\setcounter{equation}{0}
\section{Asymptotic analysis in 3D}

We now turn to the analogous problem of 3D particle clustering in the presence of anchoring sites that is maintained out-of-equilibrium by the activation/deactivation of diffusing particles; particles can only aggregate in the activated state, see Fig. \ref{fig4}. We begin by considering the mean field approximation given by Eq. (\ref{mft}) with $c(R)=0$ and 
 the flux $J(R)$ into the surface of each cluster determined self-consistently from the equation
\begin{equation}
\label{coJ3D}
J(R)=4\pi R^2 D\partial_rc(R).
\end{equation}
The solution of Eq. (\ref{mft}) in the 3D case is given by
\begin{equation}
c(r)=\left (c_0-\frac{\phi_0J(R)}{\kappa_0}\right )\left (1-\frac{R}{r}\frac{\e^{-r/\lambda}}{\e^{-R/\lambda}}\right ) .
\end{equation}
Substituting the solution into Eq. (\ref{coJ3D}) yields a self consistency equation for the flux $J(R)$:
\begin{align}
\label{mftJ3D}
J(R)=\frac{4\pi R^2D}{\lambda} \left (c_0-\frac{\phi_0J(R)}{\kappa_0}\right ) \left [1+\frac{\lambda}{R}\right ].
\end{align}
Eq. (\ref{mftJ3D}) can be rewritten as an implicit equation for the cluster radius $R$, since Eq. (\ref{SScon}) implies that $J(R)=4\pi R^3 \kappa_0 u_0/3$ \cite{Ranft23}:
\begin{align}
\label{mftR3D}
\frac{R}{\lambda}=3 \left (\frac{c_0}{u_0}-\frac{4\pi}{3}\left ( \frac{R}{\lambda}\right )^3 \phi_0\lambda^3\right )\left [1+\frac{\lambda}{R}\right ].
\end{align}
Finally, suppose that $R\ll \lambda$ and $4\phi_0 \pi \lambda^3/3 \ll 1$. Setting $R/\lambda=\epsilon \rho/\lambda$, we find that 
\begin{align}
\left (\frac{\rho}{\lambda}\right )^2 &= \frac{3c_0}{\epsilon^2 u_0}\left [1+\frac{\epsilon \rho}{\lambda}\right ].
\end{align}
The rescaled radius is $O(1)$ provided that we set $u_0=\overline{u}_0/\epsilon^3$ and $I_0=\overline{I}_0/\epsilon$. Hence,
\begin{align}
\rho &\approx  \rho_0 \left [1+\frac{\epsilon \rho_0}{2\lambda}\right ],
\label{mftX3D}
\end{align}
where
\begin{equation}
\rho_0\equiv \sqrt{\frac{3  \lambda^2\overline{c}_0}{\overline{u}_0}}.
\end{equation}
As in the 2D case, we want to determine corrections to the mean field result in the small cluster limit.

The analysis in 3D is based on an asymptotic expansion with respect to $\epsilon$ with $I_0=\overline{I}_0/\epsilon$. The outer solution is expanded as
\begin{equation}
c(\x)\sim\frac{\overline{c}_0}{\epsilon}+c_1(\x)+\epsilon c_2(\x)+\ldots
\end{equation}
with $\overline{c}_0=\overline{I}_0/\kappa_0$, and
\begin{subequations}
\label{3Douter0}
\begin{align}
D{\bm \nabla}^2 c_{\ell}(\x)-\kappa_0c_{\ell}(\x) =&0,\quad \x\in \Omega\backslash \{\x_1,\ldots,\x_N\}, \\ {\bm \nabla}c_{\ell}(\x)\cdot \n&=0,\quad \x \in \partial \Omega,
\end{align}
\end{subequations}
together with certain singularity conditions as $\x\rightarrow \x_j$, $j=1,\ldots,N$. The latter are determined by matching to the inner solution. In the inner region around the $j$-th cluster, we again introduce the stretched coordinates ${\bf y}=\epsilon^{-1}(\x-\x_j)$ and set $C_j({\bf y}) =c(\x_j+\epsilon \y)$. Expanding the inner solution as 
\begin{equation}
C_j(\y) =\frac{C_{j,0}(\y)}{\epsilon}+C_{j,1}(\y)+\ldots,
\end{equation}
we find that
\begin{align}
\label{inner3D}
{\bm \nabla}_{\bf y}^2 C_{j,\ell}&=0, \ \ell=0,1,\ {\bm \nabla}_{\bf y}^2 C_{j,2}=\overline{I}_0,\\  {\bm \nabla}_{\bf y}^2 C_{j,\ell}&=\kappa_0 C_{j,\ell-2},\ \ell \geq 3\nonumber 
\end{align}
for $ \y\in \R^3\backslash \calU_j$ and  $C_{j,\ell}(\y)=0$ for $\y \in \partial \calU_j$.
Finally, the matching condition is that the near-field behavior of the outer solution as $\x\rightarrow \x_j$ should agree with the far-field behavior of the inner solution as $|\y|\rightarrow \infty$, which is expressed as 
\[  \frac{\overline{c}_0}{\epsilon}+ c_1+\epsilon c_2 \sim \frac{C_{j,0}}{\epsilon}+  C_{j,1}+\ldots
\]

First $C_{j,0} \sim \overline{c}_0$ as $|\y|\rightarrow \infty$ so that we can set 
$ C_{j,0}(\y) =\overline{c}_0(1-w(\y))$, with $w(\y)$ satisfying the boundary value problem
\begin{align}
\label{w}
{\bm \nabla}_{\bf y}^2 w(\y)&=0,\  \y\in \R^3\backslash \calU_j ; \quad w(\y)=1,\ \y \in \partial \calU_j,\\
w(\y)&\rightarrow 0\quad \mbox{as } |\y|\rightarrow \infty.\nonumber
\end{align}
This is a well-known problem in electrostatics and has the far-field behavior
\begin{equation}
w(\y)\sim \frac{{\mathcal C}_j}{|\y|}+\frac{{\bf P}_j\cdot \y}{|\y|^3}+\ldots \mbox{as } |\y|\rightarrow \infty,
\end{equation}
where ${\mathcal C}_j$ is the capacitance and ${\bf P}_j$ the dipole vector of an equivalent charged conductor with the shape $\calU_j$. (Here ${\mathcal C}_j$ has units of length. In the case of a sphere of radius $\rho_j$ the capacitance is ${\mathcal C}_j=\rho_j$.) It now follows that $c_1$ satisfies Eq. (\ref{3Douter0}) together with the singularity condition
\[c_1(\x,s)\sim - \frac{{\overline{c}_0 \mathcal C}_j}{|\x-\x_j|} \quad \mbox{as } \x\rightarrow \x_j.\]
It follows that the solution can be written as
\begin{equation}
c_1(\x)=- {4\pi}\overline{c}_0 D \sum_{k=1}^N{\mathcal C}_kG(\x,\x_k),
\end{equation}
where $G$ is the 3D version of the Green's function defined in Eq. (\ref{G1}). Note that the 3D Green's function has the singularity structure
\begin{equation}
G(\x,\x')=\frac{1}{4\pi D|\x-\x'|} +R(\x,\x').
\end{equation}

Next we match $C_{j,1}$ with the near-field behavior of $c_1(\x)$ around the $j$-th cluster, which takes the form
\begin{align}
c_1(\x)&\sim - \frac{\overline{c}_0{\mathcal C}_j}{|\x-\x_j|}- \overline{c}_0\chi_j,
\end{align}
with
\begin{align}
\label{chi3D}
\chi_j&=4\pi D  {\mathcal C}_j R(\x_j,\x_j)+ 4\pi D \sum_{k\neq j}^N{\mathcal C}_k G(\x_j,\x_k).
\end{align}
It follows that the solution of Eq.(\ref{inner3D}) for $\ell=1$ is 
\begin{equation}
C_{j,1}(\y)=-\overline{c}_0 \chi_ j( 1-w(\y)),\  
\end{equation}
with $w(\y)$ given by Eq. (\ref{w}). Hence, $c_2$ satisfies Eq. (\ref{3Douter0}) supplemented by the singularity condition
\[c_2(\x)\sim \frac{\overline{c}_0\chi_j{\mathcal C}_j}{|\x-\x_j|} \quad \mbox{as } \x\rightarrow \x_j.\]
Following along identical lines to the derivation of $c_1(\x)$, we obtain the result
\begin{equation}
c_2(\x)=4\pi D\overline{c}_0\sum_{k=1}^N\chi_k{\mathcal C}_kG(\x,\x_k).
\end{equation}
In conclusion, the outer solution takes the form
\begin{eqnarray}
c(\x)&\sim& \overline{c}_0 \bigg[\frac{1}{\epsilon}- {4\pi}  D  \sum_{k=1}^N{\mathcal C}_k(1-\epsilon \chi_k)G(\x,\x_k)+O(\epsilon^2) \bigg ],\nonumber \\
\end{eqnarray}
while the inner solution around the $j$-th cluster is
\begin{align}
C_j(\rho)\sim  \overline{c}_0 \bigg [\frac{1}{\epsilon} -  \chi_j+O(\epsilon)\bigg ]\left (1-\frac{\rho_j}{\rho}\right ).
\end{align}

The final step is to determine the unknown radii $\rho_j$, $j=1,\ldots,N$, by imposing the conditions (\ref{SScon}), which in 3D reduces to the implicit equation
\begin{equation}
4\pi \epsilon \rho_j^2  D \overline{c}_0 \bigg [\frac{1}{\epsilon} -  \chi_j+O(\epsilon) \bigg ]\frac{1}{\rho_j} = \epsilon^3 \kappa_0 {u}_0 \frac{4\pi \rho_j^3}{3 }. \end{equation}
Setting $u_0=\overline{u}_0/\epsilon^3$ then yields 
 \begin{align}
\rho_j^2&=\frac{3  \lambda^2\overline{c}_0}{\overline{u}_0}\bigg[1-\epsilon \chi_j+O(\epsilon^2)\bigg ].
\label{rhoj3D}
\end{align}
The $O(1)$ contribution recovers the mean field result (\ref{mftX3D}) whereas the $O(\epsilon)$ coefficient $\chi_j$, see Eq. (\ref{chi3D}), is the correction to mean field theory that involves diffusion-mediated interactions between the clusters. Since the capacitance ${\mathcal C}_j=\rho_j$ for spherical clusters, it follows that keeping the $O(\epsilon)$ term results in an implicit equation for the radii, which can also be solved perturbatively. Introducing the series expansion
$\rho_j=\rho_j^{(0)}+\epsilon \rho_j^{(1)}+\ldots$, we have
\begin{equation}
\label{rho03D}
\rho_j^{(0)}=\rho_0\equiv \sqrt{\frac{3  \lambda^2\overline{c}_0}{\overline{u}_0}},
\end{equation}
and
\begin{equation}
\label{rho13D}
\rho_j^{(1)} = - 2\pi \rho_0^2D \bigg (
R(\x_j,\x_j)+   \sum_{k\neq j}^NG(\x_j,\x_k)\bigg ).
\end{equation}

\begin{figure}[t!]
\centering
\includegraphics[width=6cm]{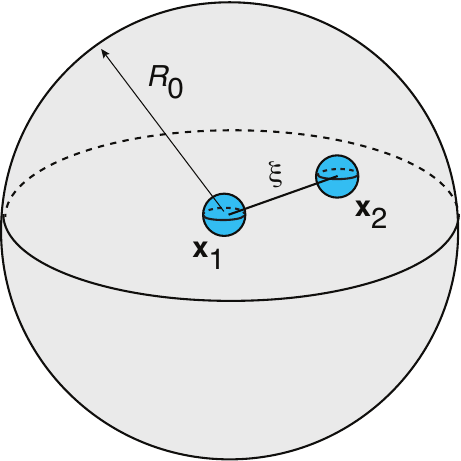} 
\caption{Pair of spherical clusters in a sphere of radius $R_0$. The first sphere is at the origin and the other is at a radial distance $\xi$ from the center.}
\label{fig5}
\end{figure}

As a simple example, consider the 3D configuration shown in Fig. \ref{fig5}. The domain $\Omega$ is taken to be a sphere of radius $R_0$ with one cluster at the origin, $\x_1={\bm 0}:=(0,0,0)$ and the other at $\x_2={\bm \xi}:= (\xi,0,0)$ with $0 < \xi <R_0$. The 3D Neumann Green's function for the modified Helmholtz equation in the sphere is known explicitly \cite{Grebenkov20}:
\begin{align}
{G} (\x,\x')&=\frac{\e^{-|\x-\x'|/\lambda}}{4\pi D|\x-\x'|}-{G}_{\rm sp}(\x,\x'),
\end{align}
with
\begin{align}
{G}_{\rm sp}(\x, \x')&=\frac{1}{4\pi D \lambda}\sum_{n=0}^{\infty} (2n+1)P_n(\cos \theta)\frac{k_n'(R_0/\lambda)}{i_n'(R_0/\lambda)}\nonumber \\
&\qquad \times i_n(|\x|/\lambda)
i_n(|\x'|/\lambda).
\label{Gsphere}
\end{align}
Here $P_n$ is a Legendre polynomial, $\x\cdot \x' =|\x||\x'|\cos \theta$, and $i_n,k_n$ are modified spherical Bessel functions,
\begin{equation}
i_n(x)=\sqrt{\frac{\pi}{2x}}I_{n+1/2}(x),\quad k_n(x)=\sqrt{\frac{2}{\pi x}}K_{n+1/2}(x).
\end{equation}
The regular part of the Green's function is
\begin{equation}
R(\x,\x)=-\frac{1}{4\pi D \lambda}-G_{\rm sp}(\x, \x).
\end{equation}
Using the identities
\begin{equation}
i_0(x)=\frac{\sinh x}{x},\quad k_0(x)=\frac{\e^{-x}}{x}, \quad i_n(0)=0, \ n>0,
\end{equation}
we see that
\begin{align}
G({\bm 0},{\bm \xi})&=\frac{1}{4\pi D \lambda}\left [\frac{e^{-\xi/\lambda}}{  \xi/\lambda}- \frac{k_0'(R_0/\lambda)}{i_0'(R_0/\lambda)}i_0(\xi/\lambda)\right ].
\label{Gsphere2}
\end{align}
Further useful identities are
\begin{align*}
(2n+1) i'_n&=ni_{n-1}+(n+1)i_{n+1},\\  -(2n+1) k'_n&=nk_{n-1}+(n+1)k_{n+1}.
\end{align*}
Eqs. (\ref{rho03D}) and (\ref{rho13D}) imply that
\begin{subequations}
\begin{align}
\rho_1 &=\rho_0+  2\pi \epsilon D  \rho_0^2\bigg (
\frac{1}{4\pi D \lambda}+G_{\rm sp}({\bm 0}, {\bm 0})-   G({\bm 0},{\bm \xi})\bigg ),\\
\rho_2 &=\rho_0+  2\pi \epsilon D  \rho_0^2\bigg (
\frac{1}{4\pi D \lambda}+G_{\rm sp}({\bm \xi}, {\bm \xi})-   G({\bm 0},{\bm \xi})\bigg ).
\end{align}
\end{subequations}
It is convenient to rewrite these equations in the form
\begin{equation}
\label{Fj}
\frac{\rho_j-\rho_0}{\rho_0}=\epsilon \frac{\rho_0}{\lambda} F_j(\xi/\lambda), \ j=1,2.
\end{equation}
The functions $F_j(\xi)$ are plotted in Fig. 6 for $R_0=5\lambda$. A number of observations can be made. (i) When the clusters are well separated, the radius of the first cluster approaches the mean-field value (\ref{mftX3D}) since $F_1(\xi)\approx  1/2$. (ii) Away from the boundary, $\xi < R_0$, the two clusters have approximately
the same radius, which is less than the mean-field value since $F_j(\xi) < 1/2$. However, as the second cluster approaches the boundary, its radius exceeds the mean-field value, that is, $F_2(\xi)>1/2$. (iii) For sufficiently small $\xi$, we have $J_j(\xi) < 0$ so that $\rho_j < \rho_0$.

\begin{figure}[t!]
\centering
\includegraphics[width=10cm]{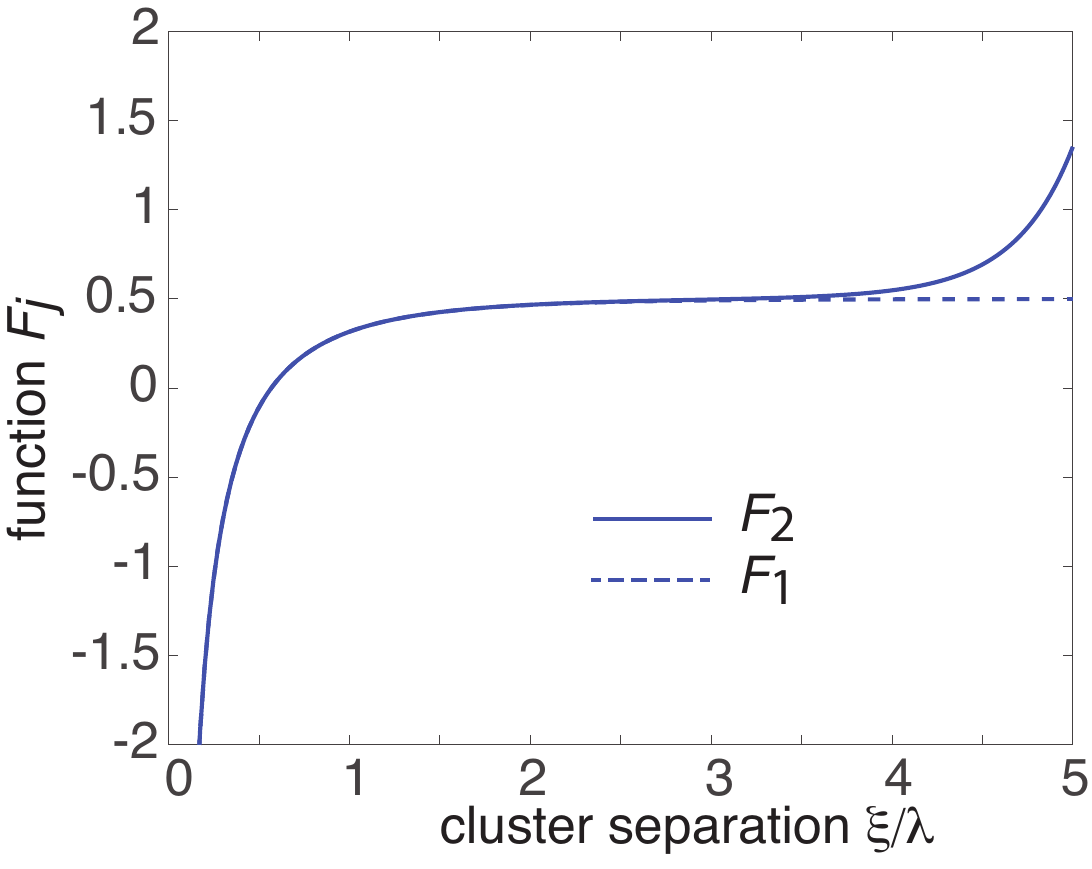} 
\caption{Plots of the multiplicative factors $F_j(\xi)$ defined in Eq. (\ref{Fj}) as a function of radial separation $\xi$ for the configuration shown in Fig. \ref{fig5}. The radius of the spherical domain $\Omega$ is $R_0=5\lambda$. }
\label{fig6}
\end{figure}

\setcounter{equation}{0}
\section{Discussion}

In this paper we used the theory of diffusion in singularly perturbed domains to calculate corrections to the mean field theory of 2D and 3D models of particle clustering in the presence of anchoring sites. We used matched asymptotics and Green's function methods to derive implicit equations for the stationary cluster sizes given by Eqs. (\ref{rhoj}) and (\ref{rhoj3D}), respectively. These equations can be solved numerically. The main result of our analysis is that diffusion-mediated interactions between the clusters generates corrections to the mean-field average radius that depend on the positions of all the anchoring sites and the exterior domain boundary $\partial \Omega$.
We illustrated our analysis in the simple case of a pair of clusters, where we explored how the radii depended on cluster separation.

One potential limitation of our steady-state analysis is to assume that a non-equilibrium steady-state solution consisting of multiple clusters exists without explicitly demonstrating that such a solution is dynamically stable. We note, however, that detailed particle-based simulations of cluster formation in the presence of anchoring sites provide strong evidence that multi-cluster states are stable, and that analytical approaches capture how cluster sizes depend on the density of anchoring sites, for example \cite{Ranft23}.

One simplification of our analysis was to ignore aggregation and cluster formation of diffusing particles
outside of the anchored domains. Such effects can be incorporated into the mean-field model by introducing a typical radius
$R_{\rm typ}$ and a typical diffusivity $D_{\rm typ}$ of diffusing
clusters. (As highlighted in Ref. \cite{Ranft23}, estimates for $D_{\rm typ}$ and $R_{\rm typ}$ could be obtained by considering stationary solutions of a set of Smoluchowski coagulation equations.) 
Eqs. (\ref{mft}) and (\ref{coJ}) are then modified accordingly \cite{Ranft23}:
\begin{align}
\label{typmft}
D_{\rm typ}{\bm \nabla}^2 c(r)-\kappa_0 c(r)+I_0-\phi_0 J(R^*)&=0
\end{align} 
for $r>R^*$, where $R^*=R+R_{\rm typ}$ is the effective radius at which diffusing and anchored clusters fuse, $c(R^*)=0$,  and (in the 2D case)
\begin{equation}
\label{typcoJ}
J(R^*):=2\pi R^* D_{\rm typ}\partial_rc(R^*)=4\pi R^2 \kappa_0\phi_0 .
\end{equation}
The asymptotic analysis would proceed as in Sect. 3 with $D\rightarrow D_{\rm typ}$, $\lambda \rightarrow \sqrt{D_{\rm typ}/\kappa_0}$, $R_j^*=\epsilon \rho_j$ and $R_j=\epsilon(\rho_j-\rho_{\rm typ})$. 

A natural generalization of the model given by Eq. (\ref{cluster}) would be to modify the absorbing boundary condition on the surface of each cluster. For example, we could replace the Dirichlet boundary condition $c(\x)=0$ by the Robin boundary condition $D{\bm \nabla}c(\x)\cdot \n_j=\kappa_0 c(\x)$ for $\x \in \partial \calU_j$, where $\kappa_0$ is the surface reactivity. (The analysis of diffusion in domains with partially absorbing traps has been analyzed elsewhere within the context of narrow capture problems \cite{Bressloff22}.)  Finally, note that in order to exploit asymptotic methods, we assumed that the clusters were well-separated, which corresponds to the dilute cluster limit of mean-field theory. In the case of a dense random or periodic distribution of clusters, it might be possible to use homogenization theory to develop an effective medium approximation of the multi-cluster system.

\end{document}